# Dilatation symmetry of the Fokker-Planck equation and anomalous diffusion

Sumiyoshi Abe

*Institute of Physics*, *University of Tsukuba, Ibaraki 305-8571*, *Japan*

Based on the canonical formalism, the dilatation symmetry is implemented to the Fokker-Planck equation for the Wigner distribution that describes atomic motion in an optical lattice. This reveals the symmetry principle underlying the recent result obtained by Lutz [Phys. Rev. A **67**, 051402(R) (2003)] on the connection between anomalous transport in the optical lattice and Tsallis statistics in the high-energy regime. Lutz's discussion is generalized to the nonstationary case, and the condition, under which the solution distribution decays as a power law, is derived.

PACS numbers: 05.60.-k, 02.20.-a, 42.50.Vk, 32.80.Pj

Anomalous diffusion [1,2] has continuously been attracting attention over decades. It can be observed, for example, in turbulent flows [3], dissolved micelles [4], chaotic dynamics [5], porous glasses [6], and subrecoil laser cooling [7].

Anomalous diffusion, or superdiffusion more specifically, may be a signal of scale-free non-Gaussian statistics. Lévy statistics [8] is one such example and is characterized by a stable class of distributions. In the one-dimensional symmetric case, the Lévy distribution indexed by $\gamma$ may be defined through the characteristic function of the stretched-exponential form: $L_\gamma(x) = (1/2\pi) \int dk \exp(ikx - a|k|^\gamma)$, where $\gamma \in (0, 2)$ and $a$ is a positive constant. Excepting the Gaussian case ($\gamma \to 2-0$), $L_\gamma(x)$ decays as a power law: $L_\gamma(x) \sim |x|^{-1-\gamma}$. An important point is the following. There exists a mathematical result termed the Lévy-Gnedenko generalized central limit theorem [9], which states that, by $N$-fold convolution, a distribution with divergent lower moments tends to one of the Lévy stable class, $\{L_\gamma(x)\}_\gamma$, in the limit $N \to \infty$ if such a limit is convergent. This is in contrast to the ordinary central limit theorem for distributions with the finite second moments in normal Gaussian statistics. Now, assume $L_\gamma(x)$ to describe a single jump in the probabilistic process. After $N$ independent jumps, one has $L_\gamma^{(N)}(x) = N^{-1/\gamma} L_\gamma(x/N^{1/\gamma})$. Identifying $N$ with time, $t$, one obtains the scaling of the distribution, $P(x,t) \equiv L_\gamma^{(N)}(x) = t^{-1/\gamma} P(x/t^{1/\gamma})$, implying that spreading of the distribution follows the law of superdiffusion, $\sim t^{1/\gamma}$, which should be compared with the case of normal diffusion, $\sim t^{1/2}$.

In a recent paper [10], Lutz has reexamined the problem of anomalous transport of atoms in an optical lattice. He has considered the following generalized Fokker-Planck equation for the marginal Wigner distribution of the momentum, $p$, of an atom in the optical potential:

$$\frac{\partial W(p,t)}{\partial t} = -\frac{\partial}{\partial p}[K(p)\, W(p,t)] + \frac{\partial}{\partial p}\left[D(p)\frac{\partial W(p,t)}{\partial p}\right]. \qquad (1)$$

Both the drift, $K$, and the diffusion coefficient, $D$, explicitly depend on the momentum of the atom as follows:

$$K(p) = -\frac{\alpha\, p}{1+(p/p_c)^2}, \qquad D(p) = D_0 + \frac{D_1}{1+(p/p_c)^2}. \qquad (2)$$

Here, $\alpha$ and $p_c$ are the damping coefficient and the capture momentum, respectively. $D_0$ and $D_1$ are constants related to fluctuations caused by the photon processes. Eq. (1) can be derived, after spatial average, from the master equation for the full Wigner distribution constructed by quantum dynamics of the atom-laser interaction (see Ref. [10] and the references therein). It has been noticed [10] that $K$ and $D$ satisfy the relation

$$\frac{K(p)}{D(p)} = \frac{d}{dp} \ln e_q(-\beta \varepsilon(p)), \tag{3}$$

where $e_q(s)$ is the "q-exponential function" defined by $e_q(s) = (1+(1-q)s)_+^{1/(1-q)}$ with the notation $(a)_+ = \max\{0, a\}$, and

$$q = 1 + \frac{2D_0}{\alpha p_c^2}, \quad \beta = \frac{\alpha}{2(D_0 + D_1)}, \quad \varepsilon(p) = p^2. \tag{4}$$

In the limit $q \to 1$, $e_q(s)$ converges to the ordinary exponential function, $e^s$. From Eq. (3), it follows that the exact stationary solution of Eq. (1) is given by [10]

$$W_q(p) = \frac{1}{Z_q(\beta)} e_q(-\beta \varepsilon(p)) \tag{5}$$

with

$$Z_q(\beta) = \int dp \, e_q(-\beta \varepsilon(p)), \tag{6}$$

which optimizes the Tsallis entropy [11,12], $S_q[W] = (1-q)^{-1} \left[ \int dp \, W^q(p) - 1 \right]$, under the constraints on normalization of $W$ and the "q-expectation value" of the "energy", $\varepsilon(p)$ [13,14]. $\beta$ is related to the Lagrange multiplier associated with the q-expectation value of $\varepsilon(p)$ (see Refs. [13,14] for more details). From these discussions, we see that the entropic index, $q$, is determined by dynamics as in Eq. (4).

It should be mentioned that since $q$ is larger than unity, the distribution in Eq. (5) decays as a power law. In fact, it has explicitly been shown [15] that if $q \in (5/3, 3)$, many-time convolution of the distribution of the form in Eq. (5) converges to the Lévy distribution with the index, $\gamma = (3-q)/(q-1)$, in accordance with the Lévy-Gnedenko generalized central limit theorem.

Now, emergence of an asymptotically power-law distribution from the linear Fokker-Planck equation is rather peculiar [16-18]. In the present case, the origin of the power law is in the behavior of the ratio in Eq. (3), that is,

$$\frac{K(p)}{D(p)} \sim \frac{1}{p}, \tag{7}$$

in the high-energy regime. One may wonder if there is an underlying principle for the emergence of this scale-free nature. In what follows, we reveal such a principle by taking advantage of the dilatation symmetry implemented to the linear Fokker-Planck equation. As a result, without assuming stationarity, we shall obtain the condition for the solution, $W$, to be asymptotically scale-free, which turns out to be more general than Eq. (7).

Our starting point is the variational principle for the kinetic equations [19-21]. The action and the Lagrangian density respectively read

$$I[W, \Lambda] = \iint dt\, dp\, \pounds\, (W, \Lambda, \partial W/\partial t, \partial \Lambda/\partial t, \partial W/\partial p, \partial \Lambda/\partial p), \qquad (8)$$

$$\pounds = \frac{1}{2}\left(\Lambda \frac{\partial W}{\partial t} - \frac{\partial \Lambda}{\partial t} W\right) - K \frac{\partial \Lambda}{\partial t} W + D \frac{\partial \Lambda}{\partial p} \frac{\partial W}{\partial p}, \qquad (9)$$

where $\Lambda(p,t)$ is an auxiliary field. Performing integration by parts and dropping the boundary terms, we see that Eq. (8) can also be expressed in the following form: $I = -\int dt\, \langle \partial \Lambda/\partial t + K \partial \Lambda/\partial p + \partial(D\partial \Lambda/\partial p)/\partial p \rangle$, where $\langle A \rangle$ stands for the ordinary expectation value of $A$: $\langle A \rangle \equiv \int dp\, A(p,t)\, W(p,t)$. Taking the variations of the action with respect to $\Lambda$ and $W$, we obtain Eq. (1) and $\partial \Lambda/\partial t + K \partial \Lambda/\partial p + \partial(D\partial \Lambda/\partial p)/\partial p = 0$, respectively.

Let us proceed to developing the canonical formalism. The canonical momenta conjugate to $W$ and $\Lambda$ are given by

$$\Pi_W = \frac{\partial \pounds}{\partial(\partial W/\partial t)} = \frac{1}{2}\Lambda, \qquad (10)$$

$$\Pi_\Lambda = \frac{\partial \pounds}{\partial(\partial \Lambda/\partial t)} = -\frac{1}{2}W, \qquad (11)$$

respectively, leading to a pair of the constraints

$$\chi_1 = \Pi_W - \frac{1}{2}\Lambda \approx 0, \tag{12}$$

$$\chi_2 = \Pi_\Lambda + \frac{1}{2}W \approx 0. \tag{13}$$

Presence of these constraints is simply due to the fact that the equations for $W$ and $\Lambda$ are of the first order in time. The basic equal-time Poisson bracket relations are

$$\{W(p,t), \Pi_W(p',t)\} = \delta(p-p'), \tag{14}$$

$$\{\Lambda(p,t), \Pi_\Lambda(p',t)\} = \delta(p-p'). \tag{15}$$

In Dirac's terminology [22], the constraints in Eqs. (12) and (13) are of the second class, since $\{\chi_1(p,t), \chi_2(p',t)\} = -\delta(p-p')$ which does not vanish weakly. Then, to eliminate these second-class constraints, it is conventional to introduce the Dirac bracket [22] defined by

$$\{A(t), B(t)\}^* = \{A(t), B(t)\} - \sum_{i,j=1}^{2} \iint dp\,dp'\,\{A(t), \chi_i(p,t)\}C_{ij}(p,p')\{\chi_j(p',t), B(t)\}, \tag{16}$$

where $A$ and $B$ are functionals of $W$ and $\Lambda$. In this equation, $C_{ij}(p, p')$ are quantities satisfying the equations: $\sum_{k=1}^{2}\int dp'' \{\chi_i(p,t), \chi_k(p'',t)\} C_{kj}(p'', p') = \delta_{ij}\delta(p-p')$.

In the present case, $C_{ij}(p, p')$ are given as follows: $C_{11}(p, p') = C_{22}(p, p') = 0$, $C_{12}(p, p') = -C_{21}(p, p') = \delta(p - p')$. Therefore, we obtain the basic relation

$$\{W(p,t), \Lambda(p',t)\}^* = \delta(p-p'), \tag{17}$$

which implies that $W$ and $\Lambda$ are canonically conjugate to each other with respect to the Dirac bracket.

The Hamiltonian is given by

$$H = \int dp \left( \Pi_W \frac{\partial W}{\partial t} + \Pi_\Lambda \frac{\partial \Lambda}{\partial t} - \pounds \right)$$

$$= \int dp \left( K \frac{\partial \Lambda}{\partial p} W - D \frac{\partial \Lambda}{\partial p} \frac{\partial W}{\partial p} \right), \tag{18}$$

and is clearly a constant of motion. Using Eqs. (17) and (18), we can ascertain that time evolution of $W$, that is,

$$\frac{\partial W(p,t)}{\partial t} = \{W(p,t), H\}^*, \tag{19}$$

precisely reproduces the Fokker-Planck equation in Eq. (1). The equation for $\Lambda$ is also described in a similar form.

Next, we consider the generator of the dilatation transformation

$$G = \int dp\, p \frac{\partial \Lambda}{\partial p} W. \tag{20}$$

This quantity gives rise to the following relations:

$$\{G(t), W(p,t)\}^* = \frac{\partial}{\partial p}[p W(p,t)], \tag{21}$$

$$\{G(t), \Lambda(p,t)\}^* = p \frac{\partial \Lambda(p,t)}{\partial p}. \tag{22}$$

Therefore, the finite transformations are expressed as

$$\exp\{(\ln\lambda)G(t)\}^* W(p,t) \exp\{-(\ln\lambda)G(t)\}^*$$
$$= e^{(\ln\lambda)\frac{\partial}{\partial p}p} W(p,t) = \lambda W(\lambda p, t), \tag{23}$$

$$\exp\{(\ln\lambda)G(t)\}^* \Lambda(p,t) \exp\{-(\ln\lambda)G(t)\}^*$$

$$= e^{(\ln\lambda)p\frac{\partial}{\partial p}} \Lambda(p,t) = \Lambda(\lambda p, t), \qquad (24)$$

where $\lambda$ is a positive constant and $e^{\{A\}^*} B e^{-\{A\}^*} \equiv B + \{A,B\}^* + (1/2!)\{A,\{A,B\}^*\}^* + \cdots$. Normalization of $W$ is kept unchanged, whereas the auxiliary field, $\Lambda$, need not be normalized. The Dirac bracket relation in Eq. (17) is preserved, as it should be, since the transformations are canonical.

Now, we are at the position to discuss the dilatation symmetry of the system. Such a symmetry is characterized by the equation

$$\{G, H\}^* = 0. \qquad (25)$$

This invariance principle may tell us under what conditions the Fokker-Planck equation in Eq. (1) admits a scale-free solution. After some calculations using Eqs. (18), (21), and (22), we obtain

$$\{G, H\}^*$$

$$= \int dp \frac{\partial \Lambda}{\partial p} \left[ KW - p\frac{\partial K}{\partial p}W - D\frac{\partial W}{\partial p} + p\frac{\partial}{\partial p}\left(D\frac{\partial W}{\partial p}\right) - D\frac{\partial}{\partial p}\left(p\frac{\partial W}{\partial p}\right)\right]. \qquad (26)$$

Therefore, invariance under the dilatation transformation gives rise to the condition

$$KW - p\frac{\partial K}{\partial p}W - D\frac{\partial W}{\partial p} + p\frac{\partial}{\partial p}\left(D\frac{\partial W}{\partial p}\right) - D\frac{\partial}{\partial p}\left(p\frac{\partial W}{\partial p}\right) = 0. \tag{27}$$

We note that here $W$ need not be stationary. We rewrite Eq. (27) as

$$\frac{\partial}{\partial p}\ln W = \frac{1}{p}\frac{\frac{\partial}{\partial p}\left(\frac{K}{p}\right)}{\frac{\partial}{\partial p}\left(\frac{D}{p^2}\right)}. \tag{28}$$

The asymptotic power-law behavior of $W$ means that

$$W(p, t) \sim \frac{a(t)}{p^\sigma} \tag{29}$$

holds for large values of $p$, where $a(t)$ is a positive function of time and the exponent, $\sigma$, is assumed to be a positive constant independent of time. Then, $\partial \ln W / \partial p \sim -\sigma / p$, and Eq. (28) gives the condition

$$\frac{K(p)}{p} + \sigma\frac{D(p)}{p^2} = c, \tag{30}$$

where $c$ is a constant. Eq. (30) is our main result. Clearly, Eq. (7) can satisfy Eq. (30) in some circumstances including Eq. (2), but the latter is more general than the former. We again emphasize that Eq. (30) is free from the assumption of $W$ to be stationary.

To see the meaning of $c$ in Eq. (30), it is necessary to consider the law of time evolution, i.e., the Fokker-Planck equation. Let us take Eq. (1) without assuming Eq. (2). Substituting Eq. (29) into Eq. (1) and using Eq. (30), we find

$$\frac{d a(t)}{d t} \sim c(\sigma - 1) a(t), \tag{31}$$

or its solution

$$a(t) \sim a(0) \, e^{c(\sigma-1)t}. \tag{32}$$

Thus, we see that $c$ in Eq. (30) is related to the asymptotic factor in Eq. (29) as follows:

$$c \sim \frac{1}{\sigma - 1} \ln \frac{a(t)}{a(0)}. \tag{33}$$

In conclusion, we have developed a general method to assess the asymptotic scale-free nature of the solution of the Fokker-Planck equation. We have seen, in the special case when the distribution is stationary, how this method can reveal the symmetry principle underlying Lutz's result on the connection between anomalous transport in the optical lattice and Tsallis statistics in the high-energy regime. We have also derived the condition in the general nonstationary case, under which the solution distribution becomes scale-free.

In the present study, we have treated only the linear Fokker-Planck equation. It has been shown [23] that a certain kind of nonlinear generalizations of the Fokker-Planck equation, which appear in various physical situations, can also yield solutions that decay as a power law. Applications of the present method to such generalized equations and systems will be discussed elsewhere.

This work was supported in part by the Grant-in-Aid for Scientific Research of Japan Society for the Promotion of Science.